\documentclass[aps,pre,groupedaddress,amssymb,amsmath]{revtex4}
\usepackage[dvips]{graphicx}

\begin{document}
\title{An Explicit Microreversibility Violating Thermodynamic Markov Process}
\author{Michael J.\ Lee}
\affiliation{Department of Physics and Astronomy, University of Canterbury, Christchurch, New Zealand}
\begin{abstract}
We explicitly construct a non-microreversible transition matrix for a Markov process and apply it to the standard three-state Potts model.
This provides a clear and simple demonstration that the usual micoreversibility property of thermodynamical Monte Carlo algorithms is not strictly necessary from a mathemetical point of view.
\end{abstract}
\maketitle

The practice of statistical physics often involves Monte Carlo studies of model thermodynamic systems such as the Ising spin-lattice \cite{1925ising,1944onsager,1989zamolodchikov} and the Potts model \cite{1952potts,1982wu}.
These studies are normally conducted with a Markov process that takes a random walk through the space of all allowed system configurations, producing a chain of states in which each individual state appears with a frequency proportional to its Gibbs probability \cite{gould-tobochnik-3}.
Such a Markov process must satisfy two conditions.
The first of these is the accessibility criterion, that from any given initial state it must be possible to evolve the system, given sufficient time, into any other possible state.
This is equivalent to the physical property of ergodicity.
The second requirement, in order for the equilibrium Gibbs ensemble to exist, is that the state-to-state transition probabilities of the Markov process must satisfy the invariance condition
\begin{equation}
\label{e-invariance}
\forall i\mathrm{,} \quad \pi_i = \sum_{\mathrm{all\ }j}\pi_jp_{ji}\mathrm{.}
\end{equation}
It is much more common, however, to see algorithms for modeling physical systems that are based upon the stronger condition of microreversibility (or detailed balance)
\begin{equation}
\label{e-microreversibility}
\pi_i p_{ij} = \pi_j p_{ji} \quad \forall i,j\mathrm{.}
\end{equation}
Such algorithms include those of Metropolis \emph{et al.\ }\cite{1953metropolis}, Barker \cite{1965barker}, and Hastings \cite{1970hastings}.
Some texts will even state that equation (\ref{e-microreversibility}) is a requirement for any Markov process in which the frequency of occurrence of a given state in the chain is proportional to its Gibbs probability \cite{binney-dowrick-fisher-newman}.
Hastings has noted that this is not the case \cite{1970hastings} and Handscomb has constructed a non-microreversible Markov process for sub-systems of three states of equal Gibbs probability \cite{1962handscomb}.
Here we shall construct a non-microreversible Markov process for a general three-state system and apply it to the standard Potts model.
Note that for a system of only two states, such as the Ising model, the microreversibility and invariance conditions are equivalent.

A Markov process can be written in matrix form
\begin{equation}
\label{e-markov}
\boldsymbol{\pi} = \mathsf{M}\boldsymbol{\pi}
\end{equation}
where $\boldsymbol{\pi}$ is the vector of Gibbs probabilities $\pi_{i}$ for all possible states $i$, and the transition matrix $\mathsf{M}$ has the set of all state-to-state transition probabilities $p_{ij}$ for its elements.
Observe that this expression is equivalent to the invariance condition, equation (\ref{e-invariance}), and so this equation must hold true if a thermodynamical equilibrium ensemble is to exist.
Because we must observe something when sampling the system, we require both
\begin{equation}
\pi_{i} \in [0,1] \quad \forall i
\end{equation}
and completeness;
\begin{equation}
\sum_{\mathrm{all\ }i} \pi_i = 1\mathrm{.}
\end{equation}
Since something well defined must take place at each step of the Markov process, we also require
\begin{equation}
\label{e-positiveelements}
p_{ij} \in [0,1] \quad \forall i,j
\end{equation}
and that the outgoing transition probabilities for each state must sum to unity;
\begin{equation}
\label{e-elementssum}
\sum_{\mathrm{all\ }j} p_{ij} = 1 \quad \forall i\mathrm{.}
\end{equation}

Now, consider the Markov process transition matrix for a three state system
\begin{equation}
\mathsf{M} = \left( \begin{array}{ccc}
p_\mathrm{aa} & p_\mathrm{ba} & p_\mathrm{ca} \\
p_\mathrm{ab} & p_\mathrm{bb} & p_\mathrm{cb} \\
p_\mathrm{ac} & p_\mathrm{bc} & p_\mathrm{cc} \\
\end{array} \right)\mathrm{.}
\end{equation}
Without loss of generality, we order the states such that $\pi_\mathrm{a} \leq \pi_\mathrm{b} \leq \pi_\mathrm{c}$.
This leaves us free to choose $p_\mathrm{ac} = (1-\alpha)$, where $\alpha \in [0,1]$.
We then have constrained choice of $p_\mathrm{ab} = \alpha(1-\beta)$, where $\beta \in [0,1]$.
Equation (\ref{e-elementssum}) implies $p_\mathrm{aa} = \alpha\beta$.
We can choose $p_\mathrm{ca} = (1-\alpha\beta)\gamma(\pi_\mathrm{a}/\pi_\mathrm{c})$, where $\gamma \in [0,1]$.
Invariance requires that $p_{ba} = (1-\alpha\beta)(1-\gamma)(\pi_\mathrm{a}/\pi_\mathrm{b})$.
A fourth free choice is $p_\mathrm{bc} = \theta$, from which it follows, by equations (\ref{e-invariance}) and (\ref{e-elementssum}), that $p_\mathrm{bb} = 1-\theta-(1-\alpha\beta)(1-\gamma)(\pi_\mathrm{a}/\pi_\mathrm{b})$, $p_\mathrm{cb} = (1-\alpha-(1-\alpha\beta)\gamma)(\pi_\mathrm{a}/\pi_\mathrm{c}) + \theta(\pi_\mathrm{b}/\pi_\mathrm{c})$, and $p_\mathrm{cc} = 1 - (1-\alpha)(\pi_\mathrm{a}/\pi_\mathrm{c}) - \theta(\pi_\mathrm{b}/\pi_\mathrm{c})$.
The constraint of equation (\ref{e-positiveelements}) requires
\begin{equation}
\theta \in \left[0, \min\left\{\frac{\pi_\mathrm{c}-(1-\alpha)\pi_\mathrm{a}}{\pi_\mathrm{b}},1-(1-\alpha\beta)(1-\gamma)\frac{\pi_\mathrm{a}}{\pi_\mathrm{b}}\right\}\right]\mathrm{.}
\end{equation}
so that writing
\begin{equation}
\label{e-thetalimits}
\theta = \delta \min\left\{\frac{\pi_\mathrm{c}-(1-\alpha)\pi_\mathrm{a}}{\pi_\mathrm{b}},1-(1-\alpha\beta)(1-\gamma)\frac{\pi_\mathrm{a}}{\pi_\mathrm{b}}\right\}
\end{equation}
yields a matrix $\mathsf{M}$ with four independent free parameters $\alpha$, $\beta$, $\gamma$ and $\delta$ all chosen from the interval $[0,1]$.
The right hand term in the brackets of equation (\ref{e-thetalimits}) is guaranteed to be the lesser of the two if $\pi_\mathrm{c} > 1/2$, or if $\alpha-\gamma > \alpha\beta(1-\gamma)$.
Otherwise the upper bound of $\theta$ will have to be determined on a case-by-case basis.
This results in a somewhat unsatisfactory algebraic expression for $\mathsf{M}$, that can be avoided through a restriction to $\delta = 0$.
This yields a transition matrix with three independent parameters and no complications;
\begin{equation}
\label{e-deltazeromatrix}
\mathsf{M} = \left( \begin{array}{ccc}
\alpha\beta & (1-\alpha\beta)(1-\gamma)\frac{\pi_\mathrm{a}}{\pi_\mathrm{b}} & \gamma(1-\alpha\beta)\frac{\pi_\mathrm{a}}{\pi_\mathrm{c}} \\
\alpha(1-\beta) & 1-(1-\alpha\beta)(1-\gamma)\frac{\pi_\mathrm{a}}{\pi_\mathrm{b}} & (1-\alpha-\gamma(1-\alpha\beta))\frac{\pi_\mathrm{a}}{\pi_\mathrm{c}} \\
(1-\alpha) & 0 & 1-(1-\alpha)\frac{\pi_\mathrm{a}}{\pi_\mathrm{c}} \\
\end{array} \right)\mathrm{.}
\end{equation}
Further choosing $\alpha = 1$ and $\gamma = 0$ reduces this to a two state system.
Taking $\beta = 0$ gives the Metropolis algorithm \cite{1953metropolis}
\begin{equation}
\left\lgroup \begin{array}{c} \pi_\mathrm{a} \\ \pi_\mathrm{b} \\ \end{array} \right\rgroup
= 
\left( \begin{array}{cc}
0 & \frac{\pi_\mathrm{a}}{\pi_\mathrm{b}} \\
1 & 1 - \frac{\pi_\mathrm{a}}{\pi_\mathrm{b}} \\
\end{array} \right)
\left\lgroup \begin{array}{c} \pi_\mathrm{a} \\ \pi_\mathrm{b} \\ \end{array} \right\rgroup\mathrm{,}
\end{equation}
setting $\beta = \pi_\mathrm{a}/(\pi_\mathrm{a}+\pi_\mathrm{b})$, gives Barker's algorithm \cite{1965barker}
\begin{equation}
\left\lgroup \begin{array}{c} \pi_\mathrm{a} \\ \pi_\mathrm{b} \\ \end{array} \right\rgroup
= 
\left( \begin{array}{cc}
\frac{\pi_\mathrm{a}}{\pi_\mathrm{a}+\pi_\mathrm{b}} & \frac{\pi_\mathrm{a}}{\pi_\mathrm{a}+\pi_\mathrm{b}} \\
\frac{\pi_\mathrm{b}}{\pi_\mathrm{a}+\pi_\mathrm{b}} & \frac{\pi_\mathrm{b}}{\pi_\mathrm{a}+\pi_\mathrm{b}} \\
\end{array} \right)
\left\lgroup \begin{array}{c} \pi_\mathrm{a} \\ \pi_\mathrm{b} \\ \end{array} \right\rgroup\mathrm{,}
\end{equation}
and $\beta = 1-\xi \pi_\mathrm{b}/(\pi_\mathrm{a}+\pi_\mathrm{b})$, with $\xi \in \left[0,(\pi_\mathrm{a}+\pi_\mathrm{b})/\pi_\mathrm{b}\right]$, gives Hasting's algorithm \cite{1970hastings}
\begin{equation}
\left\lgroup \begin{array}{c} \pi_\mathrm{a} \\ \pi_\mathrm{b} \\ \end{array} \right\rgroup
= 
\left( \begin{array}{cc}
1-\xi\frac{\pi_\mathrm{b}}{\pi_\mathrm{a}+\pi_\mathrm{b}} & \xi\frac{\pi_\mathrm{a}}{\pi_\mathrm{a}+\pi_\mathrm{b}} \\
\xi\frac{\pi_\mathrm{b}}{\pi_\mathrm{a}+\pi_\mathrm{b}} & 1-\xi\frac{\pi_\mathrm{a}}{\pi_\mathrm{a}+\pi_\mathrm{b}} \\
\end{array} \right)
\left\lgroup \begin{array}{c} \pi_\mathrm{a} \\ \pi_\mathrm{b} \\ \end{array} \right\rgroup\mathrm{.}
\end{equation}
This, of course, reduces to Barker's algorithm if $\xi = 1$, and to the Metropolis algorithm if $\xi = (\pi_\mathrm{a}+\pi_\mathrm{b})/\pi_\mathrm{b}$.

\begin{figure}[tbp]
\begin{center}
\includegraphics[width=8cm]{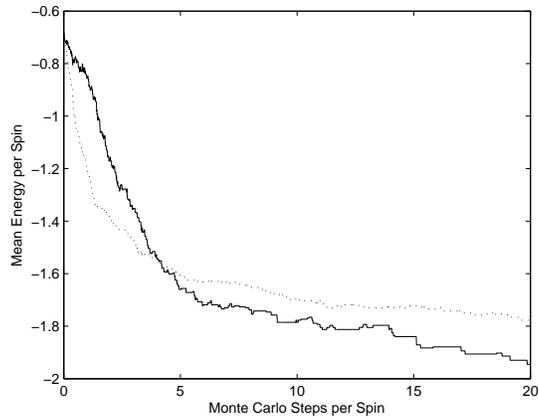}
\caption{Approach of a three-state Potts model to thermal equilibrium at $T < T_\mathrm{c}$ by Barker's algorithm (dotted line) and by the non-microreversible process of equation (\ref{e-deltazeromatrix}) with $\alpha = 0.1$, $\beta = 0.8$, and $\gamma = 0.2$ (solid line).}
\label{f-example}
\end{center}
\end{figure}

Returning to the transition matrix of equation (\ref{e-deltazeromatrix}), we see that, in general, microreversibility is violated since, for instance, $\pi_\mathrm{a} p_\mathrm{ac}$ is not identically equal to $\pi_\mathrm{c} p_\mathrm{ca}$, regardless of the choice of $\delta$.
However, it is easy to verify that the columns sum to unity, that all matrix elements lie within $[0,1]$, and that the invariance condition of equation (\ref{e-invariance}) is satisfied.
Hence this provides a legitimate Monte Carlo method that converges to an equilibrium ensemble distribution.
Figure \ref{f-example} shows the application of this method to the three-state Potts model.
This is sufficient to demonstrate the viability and practicality of Markov processes that do not have the microreversibility property.
It would of course be interesting to search the parameter space for an optimal choice to minimise thermalisation, correlation and mixing times \cite{1991goutsias}, however this search space is large and the differences are likely to be subtle around local minima.

\end{document}